\documentclass[journal]{IEEEtran}
\usepackage{graphicx}
\usepackage{amsmath}
\usepackage{subfigure}
\usepackage{rotating}
\usepackage{multirow}
\usepackage{array}
\usepackage{rotating}
\usepackage{auto-pst-pdf}

\ifCLASSINFOpdf

\else

\fi
\begin{document}
%
\title{Energy Efficient MAC Protocols}

\author{S. Hayat, N. Javaid, Z. A. Khan$^{\S}$, A. Shareef, A. Mahmood, S. H. Bouk\\\
        Department of Electrical Engineering, COMSATS\\ Institute of
        Information Technology, Islamabad, Pakistan \\
        $^{\S}$Faculty of Engineering, Dalhousie University, Halifax, Canada
        }

\maketitle

\begin{abstract}
This paper presents a survey of energy efficiency of Medium Access Control (MAC) protocols for Wireless Body Area Sensor Networks (WBASNs). We highlight the features of MAC protocols along with their advantages and limitations in context of WBASNs. Comparison of Low Power Listening (LPL), Scheduled Contention and Time Division Multiple Access (TDMA) is also elaborated. MAC protocols with respect to different approaches and techniques which are used for energy minimization, traffic control mechanisms for collision avoidance are discussed.We also present a survey of path loss models for In-body, On-body and Off-body communications in WBASNs and analytically discuss that path loss is maximum in In-body communication because of low energy levels to take care of tissues and organs located inside the body. Survey of Power model for WBANs of CSMA/CA and beacon mode is also presented.
\end{abstract}


\IEEEpeerreviewmaketitle
\section{Introduction}
\IEEEPARstart{E}{volution} of wireless, medical and computer networking technology has merged into an emerging horizon of science and technology called Wireless Body Area Networks (WBANs). However, applications of WBANs are not limited to medical field only. Miniaturization and connectivity are notable parameters of this field. WBANs consist of three levels; first level is low power sensors or nodes which are battery powered and need to be operated for a long time without repairing and maintenance. These nodes may be placed on the body, around the body or implanted in the body. Second level is called master node, gateway or coordinator which controls its child nodes; its power requirements may be less strengthened than nodes due to its applications and flexibility. Third level is the local or metropolitan or internet network that serves for monitoring purposes.

Energy efficiency or effective power consumption of a system is one of the basic requirements for WBANs because of limited power of batteries. The most suitable layer for discussing energy and power issues is MAC Layer. The basic way of saving power or enhancing energy efficiency is to minimize the energy wastage. There are several sources of energy wastage including packet collisions, over hearing, idle listening, control packet overhead, etc. Major source of energy inefficiency among the above listed sources is packet collision for WBANs. Fig. 1. best explains that how a node's battery is consumed, in the process of communication.

Collision avoidance for energy efficiency, minimum latency, high throughput, and communication reliability, are basic requirements in the design of MAC protocol. The fundamental way of saving power or enhancing energy efficiency is to minimize the energy wastage. Simulations are performed in MATLAB for different scenarios to compute path loss. Results show that path loss is maximum in In-body communication, as compared to On-body and Off-body communication because human body is composed of tissues and organs in which communication is difficult and thus results in high path loss. On-body and Off-body also results some variations when the source and destination sensors or nodes are placed Line of Sight (LoS) and Non Line of Sight (NLoS).

In this paper, we therefore, provide a survey of energy efficient  MAC protocols for WBANs.  First, we elaborate the protocol features and then their advantages and limitations are discussed. Sources that contribute to the energy inefficiency in a particular protocol is also identified. Moreover, comparisons of MAC protocols in the context of WBANs are tabulated in detail.

\section{Related Work}
Gopalan \textit{et al.} [1] survey MAC protocols for WBANs along with the comparison of four protocols i.e., Energy Efficient MAC, MedMac, Low Duty Cycle MAC, and Body MAC. Some key requirements and sources of energy wastage are also discussed. They also discussed some open research issues in this survey. Still a lot of work has to be done in data link layer, network layer and cross layer design.

In [2], Shahjahan kutty \textit{et al.} discuss the design challenges for MAC protocols for WBANs. They classify data traffic for WBANs into three categories: energy minimization techniques, frame structures and network architecture. However, the comparison of protocols is not provided by them.

Sana Ullah \textit{et al.} in [3] provide relatively a comprehensive study of MAC protocols for WBANs. Comparison of the low power listening, scheduled contention and Time Division Multiple Access (TDMA) is provided. MAC requirements, frame structures and comparison of different protocols and their trade-offs are discussed in detail.

\section{Energy Minimization Techniques in MAC Protocols for WBANs}
Low power mechanisms play an important role in performance enhancement of MAC protocol for WBANs. In this section, different approaches and techniques that provide energy efficiency in MAC protocols for WBANs are discussed and compared.

Energy efficiency is an important issue because the power of nodes in WBANs is limited and long duration of operation is expected. The key concept for low power consumption is to minimize the energy consumption in the following sources: sensing, data processing and communication.

Most of the energy wastage is caused during communication process because of the collision of packets, idle listening, over hearing, over-emitting, control packet overhead and traffic fluctuations. Idle listening can be reduced through duty cycling. To reduce energy waste in order to increase network's life time and to enhance the performance of MAC protocol, different wake-up mechanisms are used.

There are three main approaches adopted for the energy saving mechanisms in MAC protocols for WBANs, which are: Low Power Listening (LPL), Scheduled Contention, and TDMA.

\begin{figure}[h!]
  \includegraphics[height=9cm, width=9cm]{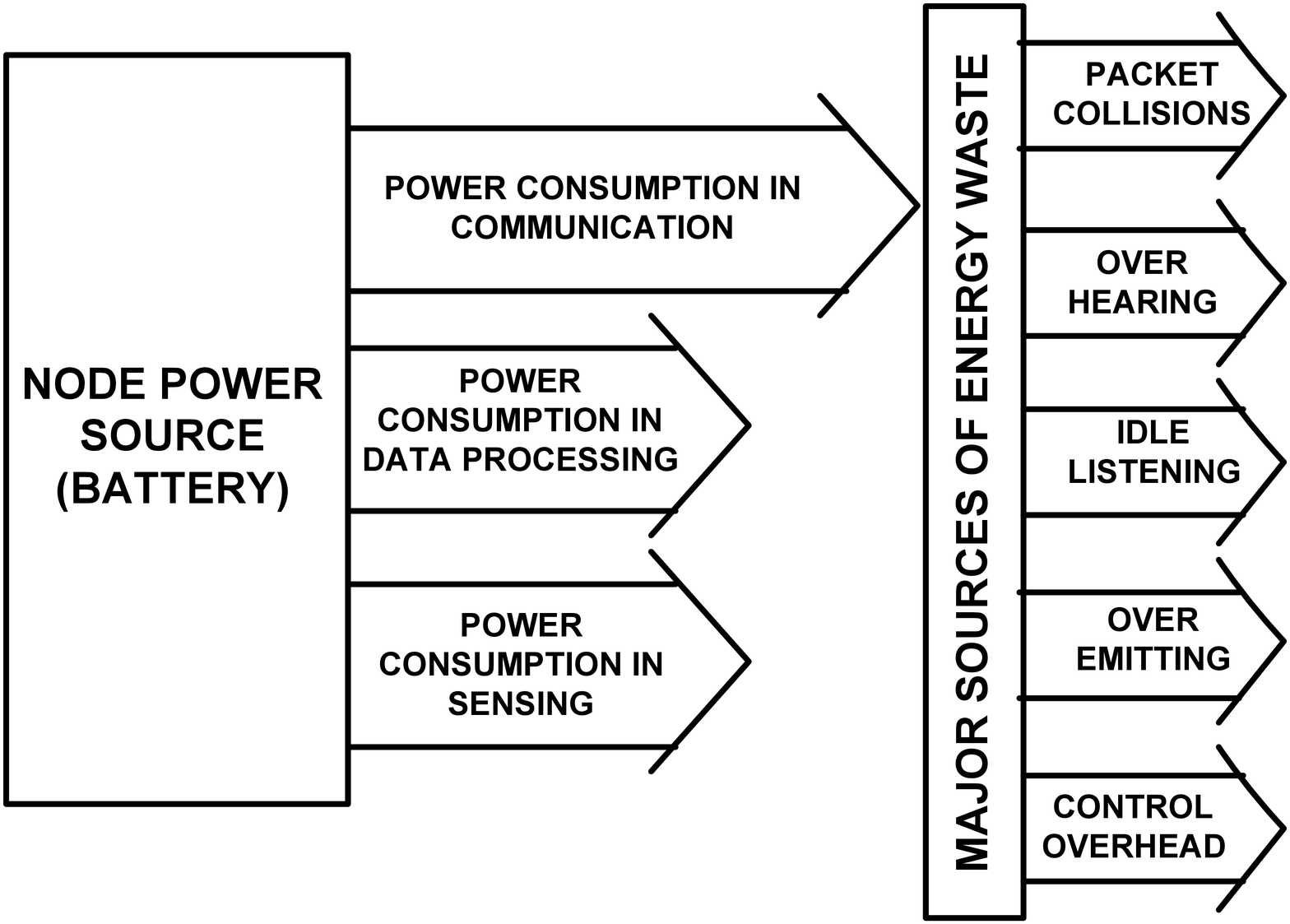}
   \caption{Sources of Power Consumption}
\end{figure}

\subsection{Low Power Listening}
LPL procedure is that ``node awakes for a very short period to check activity of channel". If the channel is not idle then the node remains in active state to receive data and other nodes go back to sleeping mode. This is also termed as channel polling [3]. This procedure is performed regularly without any synchronization among the nodes. A long preamble is used by the sender to check polling of the receiver.
LPL is sensitive to traffic rates which results in degradation of performance in the scenario of highly varying traffic rates. However, it can be optimized effectively for already known periodic traffic rates. Wise-MAC [3] is one of the MAC protocols which is based on LPL. This protocol reduce Idle listening using non-persistent CSMA and preamble sampling technique.

\subsection{Scheduled Contention}
Scheduled Contention is the combination of the scheduling and contention based mechanisms to effectively cope with the scalability and collision problems.
In contention based protocols, contending nodes try to access the channel for data transmission therefore, ability of collision of packet is greatly increased. Example of contention based MAC protocol is Carrier Sense Multiple Access/Collision Avoidance (CSMA/CA) in which Clear Channel Assessment (CCA) is performed by the nodes before transmitting data.

Scheduling or contention free means that each node has the schedule of transmission in the form of bandwidth or time slot assignment. TDMA, CDMA and FDMA schemes are some examples of scheduling mechanisms. However, CDMA and FDMA are not suitable for WBANs because of high computational overhead and frequency limitations, respectively.

TDMA is the most suitable scheduling scheme, even though it requires extra power consumption due to its sensitivity for synchronization.
The scheduled contention is the combination of scheduling and contention based mechanisms. In scheduled contention, a common schedule is adopted by all the nodes to transmit data. This schedule is exchanged periodically among the nodes to make communication adaptive, flexible and scalable.

Sensor MAC (S-MAC) is one of a MAC protocol based on the scheduled contention. In this protocol, low duty mode is set as default mode for all the nodes which assures the coordinated sleeping among neighboring nodes. The energy wastage due to collision, overhearing, idle listening etc. is minimized because the node is turned on only for transmission of data and remains in sleep mode, otherwise.

\subsection{Time Division Multiple Access }
In TDMA mechanism, a super frame consists of a fixed number of time slots is used. Time slots are allocated to the sensor nodes by a central node and is known as Master Node (MN), Cluster Head (CH), coordinator or Base Station Transceiver (BST). Traffic rate is one of the key parameter used by the coordinator to allocate time for each contending node. The scheme is power efficient because a node gets time slot for transmission of data and remains in sleep mode for rest of the time. However, the synchronization requirements may degrade performance in terms of power consumption. Therefore, it is highly sensitive to clock drift, which may result in limited throughput. Preamble-Based TDMA (PB-TDMA) protocol is one of the TDMA based protocol. Other examples include Body-MAC (B-MAC) [5], MedMAC [3] etc.

These techniques are briefly compared in Table.I.

\begin{table*}[t]
\begin{tabular} { | m {2cm}| m {5cm}| m {4cm}| m{5cm}| }
       \multicolumn{4}{c}{Table.1. Comparison between LPL, Schedule Contention, and TDMA} \\
    \hline
    \centering
   \textbf{Energy Saving Mechanisms} & \textbf{ LPL} & \textbf{Scheduled Contention} & \textbf{TDMA} \\ \hline
     Adaptability to traffic and delay & Scalable and adaptive to traffic load and low delay & Better delay performance due to sleep schedules & Better end-to-end reliability, smaller delays, high reliability \\ \hline
    Transmission latency and throughput & Flexible, high throughput, tolerable latency, and low power consumption  & High transmission latency, loosely synchronized, low throughput & Good for energy efficiency, prolonged network's lifetime, load balancing \\ \hline
    Synchronous/ Asynchronous & Asynchronous &  Synchronous  & Synchronous-Fine grained time synchronization  \\ \hline
   Traffic heterogeneity requirements &  Low duty cycle nodes do not accommodate aperiodic traffic. Very hard to satisfy the WBANs traffic heterogeneity requirements & Low duty cycle nodes do not require frequent synchronization of schedules. Hard to satisfy the WBANs traffic heterogeneity requirements & Low duty cycle nodes do not require frequent synchronization at the beginning of each superframe. Easy to satisfy the WBANs traffic heterogeneity requirements
    \\ \hline
   Sensitivity &  Sensitive to tuning for neighborhood size and traffic rate & Sensitive to clock drift & Very sensitive to clock drift \\ \hline
    Performance with respect to traffic rates&  Poor performance when traffic rates changes & With the increase in traffic, performance is improved & Throughput and number of active nodes are limited  \\ \hline
    Cost incurred by sender and receiver &  Receiver and polling efficiency is gained at much greater cost of senders & Similar cost incurred by sender and receiver & Require clustering \\ \hline
     Extravagant & It does not listen for full contention period as a result it is less expensive & Listening for full contention period  & Low duty cycle  \\ \hline
     Scalability and adaptability & Challenging to adapt LPL directly to new radios like IEEE 802.15.4 & Scalable, adaptive, and flexible & Limited scalability and adaptability to changes on number of nodes   \\ \hline

     \end{tabular}
   \end{table*}

\section{Energy Efficient MAC Protocols}
In this section, we briefly discuss the energy efficient MAC protocols for WBAN.

\subsection{Okundu MAC Protocol}
An energy efficient MAC protocol for single hop WBANs is proposed by Okundu \textit{et al.} in [4]. This protocol consists of three main processes: link establishment, wakeup service, and alarm process. Basic energy saving mechanism of this protocol consists of central control of wakeup/sleep time and Wakeup Fall-back Time (WFT) processes. WFT mechanism is used to avoid collision due to continuous time slot. This mechanism states that, if a slave node wants to communicate with a MN and it fails in its task due to MN's other activities, then it goes back to sleep mode for a specific time computed by WFT. However, data is continuously being buffered during the sleep time.

To minimize time slot collision, the concept of WFT has been introduced. This concept helps every slave node to maintain a guaranteed time slot even if it fails to communicate with the MN. In this protocol, problems like idle listening and over-hearing can be reduced because of central management of traffic.

In one cluster, only $8$ slave nodes can be connected to MN which restricts inclusion of other slave nodes. In link establishment, wakeup service, and alarm processes, communication is initiated by the MN. Another main problem is that, only one slave node can join network at a time.

\subsection{MedMac Protocol}
N. F. Timmons \textit{et al.} in [5] propose a TDMA-Based MAC protocol for WBANs called MedMAC. This protocol consists of two schemes for the power saving: Adaptive Guard Band Algorithm (AGBA) and  Drift Adjustment Factor (DAF). AGBA along with time stamp is used for synchronization among coordinator and other nodes. This synchronization is introduced using Guard Band (GB) between time slots to allow the node to sleep for many beacon periods. DAF is used to minimize bandwidth. GB is calculated by AGBA and shows the worst cases. However, practically gaps may be different between time slots depending upon application scenarios. DAF adjusts GB according to practical situation and avoids overlapping between consecutive slots.

MedMac outperforms IEEE 802.15.4 for Class 0 (lower data rate applications such as health monitoring and fitness) and Class 1 (medium data rate medical applications such as EEG). Energy waste due to collision is reduced by introducing Guaranteed Time Slot (GTS). Each device has exclusive use of a channel for a fixed time slot, therefore, synchronization overhead is also reduced.

This protocol works efficiently for low data rate applications, and work inefficiently for high data rate applications. However, In-body and On-body applications of WBAN are usually of higher data rate.

\subsection{Low Duty Cycle MAC Protocol}
Low Duty Cycle MAC protocol for WBANs is designed in [6]. In this protocol, analog to digital conversion is performed by slave nodes while the other complex tasks such as digital signal processing is carried out at MN. MNs are supposed to be less power than slave nodes.

This protocol introduces the concept of Guard Time ($Tg$) to avoid overlapping between consecutive time slots. After T frames a Network Control (NC) packet is used for general network information. Power saving is achieved by using effective TDMA strategy.

This protocol is energy efficient because it sends data in short bursts. By using TDMA strategy, this protocol effectively overcomes the collision problem. It allows monitoring patient's condition and can reduce the work load on medical staff, while keeping minimum power usage.

TDMA strategy is used in WBANs, and it is found that TDMA is more suitable for static type of networks with a limited number of sensors generating data at a fixed rate therefore, this protocol may not respond well in a dynamic topology.

\subsection{B-MAC Protocol}
B-MAC protocol achieves energy efficiency by using three bandwidth management schemes: Burst, Periodic and Adjust Bandwidth.

Burst bandwidth consists of temporary period of the bandwidth, which includes several MAC frames and recycled by the gateway (coordinator). Bandwidth is reduced to half if it does not fully utilized by the nodes, which is also informed about reduction of bandwidth.
Periodic bandwidth is a provision for a node to have access to the channel exclusively within a portion of each MAC frame or few MAC frames. It is also allocated by the gateway based on node's QoS requirements and current availability of the bandwidth [7].
Adjust bandwidth defines the amount of bandwidth to be added to or reduced from previous Periodic Bandwidth [7].

Nodes can enter into sleep mode and wake up only when they have to receive and transmit any data to the gateway, because the nodes and the gateway are synchronized in time. The time slot allocation in Contention Free Period (CFP) is collision free, which improves packet transmission and thus, saves energy.

The protocol uses CSMA/CA in the uplink frame of Contention Access Period (CAP) period, which is not reliable scheme due to its unreliable CCA and collision issues.


\subsection{Ta-MAC Protocol}
Traffic aware MAC (Ta-MAC) protocol utilizes traffic information to enable low-power communication. It introduces two wakeup mechanisms: a traffic-based wakeup mechanism, and a wakeup radio mechanism. Former mechanism accommodates normal traffic by exploiting traffic patterns of nodes whereas, later mechanism accommodates emergency and on-demand traffic by using a wakeup radio signal.

In the traffic-based wakeup mechanism, the operation of each node is based on traffic patterns. The initial traffic pattern is defined by the coordinator and can be changed later. The traffic patterns of all nodes are organized into a table called traffic-based wakeup table. In wakeup radio mechanism, a separate control channel is used to send a wakeup radio signal. The coordinator and the member node send wakeup radio signal in on-demand and emergency case.

In Ta-MAC, a node wakes up, whenever it has a packet to send/receive. Since the traffic patterns are pre-defined and known to the coordinator, it does not have to wait for resource allocation information/beacon. As a result, delay is minimized comparitive to other MAC protocols. This protocol accommodates normal, emergency and on-demand traffic in a reliable manner. To achieve energy efficiency in MAC protocol, the central coordination and resource allocation is based upon the traffic patterns  of the nodes.

As, in this protocol, the traffic pattern are defined by the coordinator, in a static topology. Therefore, it does not work efficient in dynamic topology (in dynamic topology, traffic patterns are changed frequently).

\begin{tiny}
\begin{table}
\begin{tabular} { | m{2.5cm}| m{2.5cm}| m{2.5cm}| }
    \multicolumn{3}{c}{Table. 2. Qualitative Comparison of MAC Protocols} \\
    \hline
    \centering
   \textbf{Protocols} & \textbf{Advantages} & \textbf{Disadvantages} \\ \hline
    Okundu MAC  & Minimize time slot collision, reduce idle listening and overhearing   & Only 8 slave nodes can be communicated to MN   \\ \hline
    MedMAC &  Energy waste due to collision is reduced  & Do not support high data rate applications  \\ \hline
    Low Duty Cycle & Collision problem is reduced, allows  patients' monitoring & Not suitable for dynamic type of networks \\ \hline
    B-MAC & Improves packet transmission hence saves energy & Uses CSMA/CA in the uplink frame of CAP period, which is not a reliable scheme  \\ \hline
    Ta-MAC & Accommodates normal, emergency and on-demand traffic, energy efficient, reasonable delay & Not suitable for dynamic topologies  \\ \hline
     S-MAC & High latency and time synchronization overhead may be prevented due to sleep schedules & Low throughput, overhearing and collision may cause if packet is not destined to listening node  \\ \hline
      T-MAC & Packets are sent in burst and with low latency which collectively gives better result under variable load & Suffers from sleeping problems  \\ \hline
    H-MAC & Improves BSN's energy efficiency and reduces extra energy cost & Does not support sporadic events and posseseslow spectral/bandwidth efficiency   \\ \hline
     DTDMA & Reduce packet dropping rate, less energy consumption & Does not support emergency and on-demand traffic
    \\ \hline
     \end{tabular}
   \end{table}
   \end{tiny}

\subsection{S-MAC Protocol }
S-MAC [8] protocol is proposed for WBASNs. This protocol uses fixed duty cycles to solve idle listening problem. Nodes wakeup after a specific time, as assigned by coordinator, sends data and goes back to sleep mode again. As, all the nodes are synchronized, therefore, collision can also be easily avoided. S-MAC gives considerably low latency. In this protocol, time synchronization overhead may be prevented due to sleep schedules.

Fluctuating traffics are not supported and no priority is given to the emergency traffic scenarios by S-MAC. Therefore, it is not a reliable for WBANs. Overhearing and collision may occur if the packet is not destined to the listening node.

\subsection{T-Mac Protocol }
Mihai \textit{et al.} [9] suggested Time-out MAC (T-MAC) for WBASNs. It uses flexible duty cycles for increasing energy efficiency. In T-MAC, the node wakes up after time slot assignment, sends pending messages. If there is no activation event for Time Interval (TA), the node goes back to sleep mode again. If a node sends Route To Send (RTS) and does not receive Clear To Send (CTS), then sends RTS two more times before going to sleep. To solve early sleep problem, it uses future RTS for taking priority on full buffer.

In T-MAC, packets are sent in burst, as a result delay is minimized. It also outperforms other MAC protocols under variable load. The main disadvantage in this protocol is that it suffers from sleeping problems.

\subsection{H-MAC Protocol }
Heartbeat Driven MAC (H-MAC) uses heart beat rhythm information for synchronization of nodes. This avoids the use of external clock and thus reducing the power consumption. Also guaranteed time slot (GTS) provision to each node helps to avoid collision.

H-MAC aims to improve BSNs energy efficiency by exploiting heartbeat rhythm information, instead of using periodic synchronization beacons to perform time synchronization [3].

Although, H-MAC protocol reduces extra energy cost of synchronization, however, it does not support sporadic events. Since TDMA slots are dedicated and are not traffic adaptive, H-MAC protocol encounters low spectral/bandwidth efficiency in case of low traffic. The heartbeat rhythm information varies depending on patient's condition. It may not reveal valid information for synchronization all the time [3].

\subsection{DTDMA Protocol}
Reservation based dynamic TDMA (DTDMA) protocol uses slotted ALOHA in CAP field of super frame to reduce collisions and to enhance power efficiency.

Through the adaptive allocation of the slots in a DTDMA frame, WBAN's coordinator adjusts the duty cycle adaptively with traffic load. Comparing with IEEE 802.15.4 MAC protocol, DTDMA provides more dependability in terms of lower packet dropping rate and low energy consumption especially for an end device of WBAN [3]. It does not support emergency and on-demand traffic. Furthermore, DTDMA protocol has several limitations when considered for the Medical Implant Communication Service (MICS) band. The MICS band has ten sub-channels and each sub-channel has $300$ Kbps bandwidth. DTDMA protocol can operate on one sub-channel, however, cannot operate on ten sub-channels simultaneously [3].

The main purposes of a MAC protocol are to provide energy efficiency, network stability, bandwidth utilization and reduce packet collision.

The energy minimization techniques and mechanism in MAC protocols are summarized in Table. 3.

\begin{table}
 \begin{tabular}{| m {3cm}| m{4cm}| }
    \multicolumn{2}{c}{Table. 3. Energy Minimization Techniques and Mechanisms} \\
    \hline
    \textbf{Protocol} & \textbf{Energy Efficiency Mechanism}  \\ \hline

    Okundu MAC & Wake up Fall back Time (WFT)  \\ \hline
    MedMAC  & TDMA, Adaptive Guard Band Algorithm (AGBA) and Drift Adjustment Factor (DAF)\\ \hline
    Low Duty Cycle & TDMA, concept of Guard Time ($Tg$)\\ \hline
    B-MAC  & TDMA, Bandwidth mechanism \\ \hline
    Ta-MAC  & Central coordination according  to traffic patterns of the nodes \\ \hline
    S-MAC  & Scheduled based, organized in slots and operation based on schedules \\ \hline
    T-MAC  & Have slots and operation is based on schedules \\ \hline
    H-MAC  & Heartbeat Rhythm information is used for synchronization \\ \hline
    DTDMA & TDMA based, use of slotted aloha in CAP field\\ \hline

    \end{tabular}
     \end{table}

\section{Performance Trade-offs made by MAC Protocols}
In this section, we discuss the performance of the MAC protocols they achieve and price they pay. In other words, trade-offs, the MAC protocols have to make.

\subsection{Okundu MAC Protocol}
Network's scalability is mainly application dependent, e.g., ECG can support upto maximum of 8 slave nodes because of 8 percent duty cycle. However, in practice this is 6 to allow for possible retransmissions. Therefore, we have a trade-off, for retransmission, slave nodes attached to the MN are reduced to attain scalability of network.

\subsection{MedMac Protocol}
The low data rate applications of Class $0$ medical devices include monitoring of respiration system, temperature of human body, pulse monitoring etc. Power consumed by respiration transceiver is slightly high in MedMAC protocol with respect to other protocols, while temperature and pulse node show much low power consumption, as compared to other protocols. MedMAC trade-offs power consumption of respiration for less power of other two applications.

\subsection{Low Duty Cycle MAC Protocol}
The number of extra slots needed for protocol robustness is dependent on Packet Error Rate (PER) and Packet Loss Ratio (PLR). When PER is high, it will increase PLR. However, PLR, may be reduced by using extra slots in the time frame. Therefore, this protocol can trade-offs extra slots for less PLR.

\subsection{B-MAC Protocol}
B-MAC trade-offs idle listening for a reduced time to transmit and reception of data. As, we know that reducing duty cycle increases sleep time which in turn reduces idle listening. Another trade-off is between idle listening and packet length, because this overhead dominates the energy consumption.

\subsection{Ta-MAC Protocol}
Ta-MAC uses two wakeup mechanisms one for handling data traffic and other for emergency traffic. By using these two mechanisms this protocol outperforms all other protocols in terms of power consumption because problems like idle listening, collision and overhearing are reduced. However, by sending frequent control messages to the nodes increases node's overhead, which is a trade-off. The initial traffic patterns of all the nodes are defined by the coordinator, as a result delay is also slightly increased.

\subsection{S-MAC Protocol }	
For transmission and reception of data in S-MAC, an extremely low duty cycle is used. When throughput  increases SAC's duty cycle also increases , which further increases  the  overhead of SYNChronization (SYNC) period, as a result, power consumption is increase linearly. S-MAC can trade-offs throughput for energy, also it can trade-offs energy for latency.

\subsection{T-Mac Protocol }
T-MAC uses adaptive duty cycle, implemented as a time out after the last event. At lower transmission rates, throughput increases because probability of packet loss is much less than received packet, however, the latency is increased between source and destination node.

\subsection{H-MAC Protocol}
In H-MAC a Guard Band is introduced in time slots to avoid collision by overlapping of data, however,  when time slots are completely aligned then there will be no data transmission in Guard Band, therefore, it reduces  bandwidth utilization. The coordinator of BSN then uses this GB for synchronization, by sending re-synchronization control packets, hence achieving energy efficiency. Thus  making a trade-offs between energy efficiency and  bandwidth utilization efficiency.

\subsection{DTDMA Protocol}
DTDMA is a TDMA based protocol which uses time slots for data transmission and as a result low power is consumed. However, TDMA requires synchronization between nodes and the coordinator, as a result, overhead is increased. This overhead is a trade-off for energy.

The trade-offs of each selected protocol are summarized and given in Table. IV.

\begin{table}
 \begin{tabular}{| m {3cm}| m{4cm}| }
    \multicolumn{2}{c}{Table. 4. Trade-offs Made by MAC Protocols} \\
    \hline
    \textbf{Protocol} & \textbf{Trade-Offs}  \\ \hline

    Okundu MAC &Trade-offs number of slave nodes attached to the MN are reduced for scalability of network\\ \hline
    MedMAC  &  Trade-offs idle listening for a reduced time to transmit and reception of data\\ \hline
    Low Duty Cycle & Can trade-off extra slots for less PLR\\  \hline
    B-MAC  &   Trade-off is between idle listening and packet length\\\hline
    Ta-MAC  &  Trade-off delay for low power consumption\\\hline
    S-MAC  &  Can trade-off energy for latency.\\\hline
    T-MAC  &  Can trade-off latency for high throughput\\\hline
    H-MAC  &  Trade-offs between energy efficiency and  bandwidth utilization efficiency\\\hline
    DTDMA &  Trade-offs overhead for a low power consumption\\\hline

    \end{tabular}
     \end{table}

 \section{MAC Frame Structure}
MAC frame structure consists of control portion or control packet and data portion. Control portion is responsible for the management and control messages (beacon period, request period, topology management period) to control and manage dynamic topology and varying data rate traffic. Data portion consist of two sub parts: CAP and Contention Free Period (CFP). CAP consists of CSMA/CA  while the nodes contend in CAP transmit MAC control packets. Similarly, small size data packets can also be transmitted in CAP.

In [7], the allocation of time slots is controlled by the coordinator. The coordinator arrange the duration of control and data packet on the basis of current traffic of topology that is why the slots are allocated to CFP and are collision free. In each frame, bandwidth allocation in CFP can be changed.

In [5], the GB is used to maintain synchronization among devices even if a node is sleeping for many beacon periods.

In [4][9][10][11], MAC protocols use slotted ALOHA in its frame structure to divide a slot into 4 equal mini slots. In [6], $Tg$ is introduced in its frame structure to reduce overlapping between the two following nodes.

\begin{figure*}{}
\centering
\includegraphics[height=11cm, width=14cm]{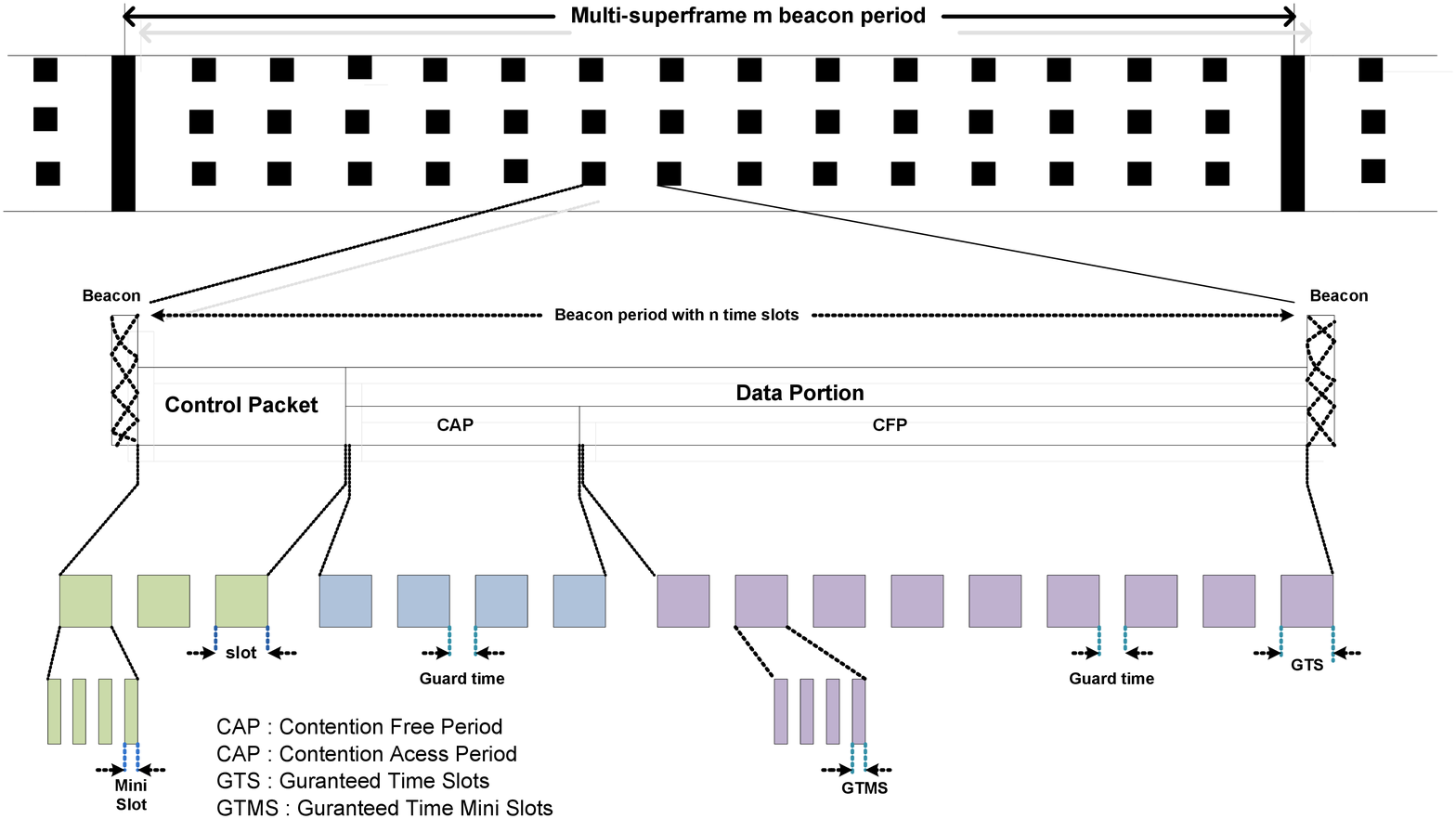}
\caption{MAC FRAME STRUCTURE}
\end{figure*}

\section{Technique for Collision Avoidance for Traffic Control}
The main schemes of MAC protocol for WBANs are divided into two groups: contention based i.e., CSMA and contention free i.e., TDMA. Most of the traffic is interrelated in WBANs, therefore, contention based solutions are not suitable for them. For example, if a patient is suffering from fever, the body temperature increases which increases blood pressure, hence, the sensor sensing temperature variation and the sensor that senses blood pressure variation, both become active. Along with them other respiration sensors also become active at the same time and try to access the channel/coordinator. However, in this situation, collision occurs in CSMA. In TDMA, each node communicate to MN according to the assigned pattern by the coordinator. As a result, collision in data traffic is low, as compared to CSMA.

\subsection{Okundu MAC }
This protocol controls traffic using centrally controlled wakeup/sleep time. Slots are assigned to sensors change every time when coordinator detects any change in traffic pattern. Assignment of different time slots, decreases collision between the nodes . It makes the system to handle fluctuating traffic. The sensor nodes establishe link with the coordinator after listening to the Radio Frequency (RF)-channel for a fixed time period. MN sends request to the sensor node for information by setting and communicating the next wakeup time after establishing the link.

\begin{table}
\centering
  \begin{tabular}{ | c| c |  }
    \multicolumn{2}{c}{Table. 5. Comparison between IEEE 802.15.4 MAC and Original IEEE 802.15.4} \\
    \hline
    \textbf{IEEE 802.15.4 MAC} & \textbf{Original IEEE 802.15.4}  \\ \hline
    Low power consumption & High power consumption  \\ \hline
    Higher data rate & Low data rate \\ \hline
    Higher flexibility & Low flexibility \\ \hline
    TDMA based & Contention based \\ \hline
    Collision Free & Greater collisions \\ \hline
    Sleep mode & Idle listening \\ \hline

    \end{tabular}
    \end{table}

\subsection{MedMAC}
MedMac reduces the collision by using AGBA. AGBA allows the sensor nodes to sleep  for a GB time period between each time slot. Each node has specific time slot to communicate with master node/coordinator, which means there is no collision. Thus minimizes the synchronization overhead.

\subsection{Ta-MAC}
Ta-MAC protocol uses two channel access mechanisms for traffic control i.e., traffic based wakeup mechanism for normal traffic, and wakeup radio mechanism for on-demand and emergency traffic. In traffic-based wakeup mechanism, all nodes have traffic patterns that are assigned by the coordinator. The initial patterns are defined and updated by the coordinator. The traffic patterns of all nodes are synchronized and arranged in a specific table, known as $ Traffic $ $ Based $ $ Wakeup $ $ Table $. Node's ID and its respective traffic patterns are stored in this table.

Normally, all the nodes become active/wakeup according to their traffic patterns. If two or more nodes have same wakeup pattern then the node with high priority is treated first by the coordinator, as shown in Fig. 3. By assigning these patterns, load at the coordinator is minimized, and chances of collision is also reduced.

\subsection{B-MAC}
B-MAC uses downlink and uplink schemes along with sleeping mode for data traffic control. Downlink is only used by MN, therefore, traffic and data load on downlink are reduced. Uplink is divided into CAP and CFP. MN allocates time slots to CFP according to data traffic which makes CFP collision free. In case, when nodes have no data to transmit or receive then they go to sleeping mode.

 \begin{figure*}[!t]
  \begin{center}
  \includegraphics[width=16cm, height=12cm]{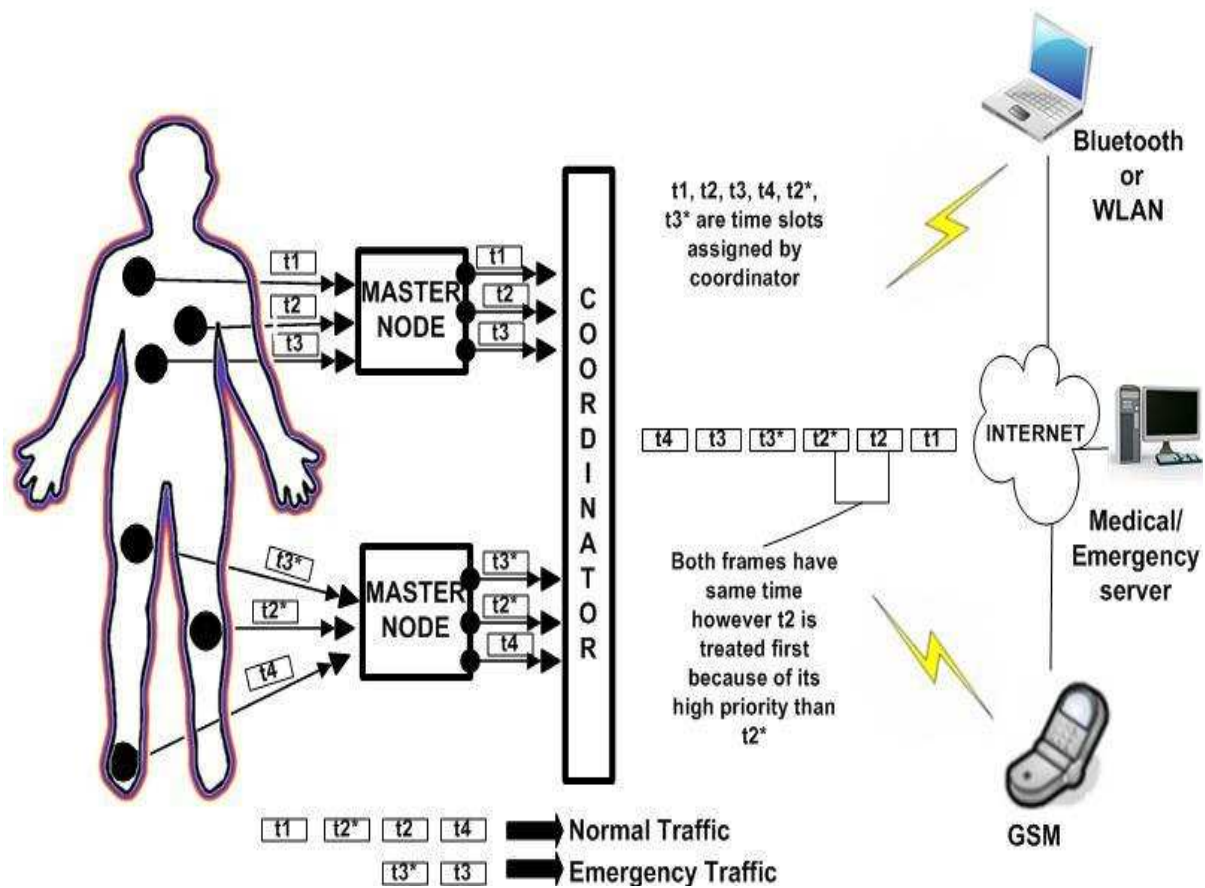}
  \caption{Data Traffic Control}
    \end{center}
\end{figure*}

\subsection{Low Duty Cycle}
Low duty cycle MAC protocol is based on TDMA. In TDMA, time slots are assigned to the sensor nodes by the coordinator. To avoid collision between the data traffic, the concept of $Tg$ is introduced. Use of $Tg$ between every consecutive slots prevents the transmission overlaps and controls data traffic.

\subsection{IEEE 802.15.4 MAC}
The basic requirement of QoS is to minimize delay and maximize the probability of successful transmission. CFP scheme is used to control data traffic to guarantee the QoS. If a node wants to send data, first it listens for the network beacon. After node finds the beacon that is sent by the coordinator, the node synchronizes to the super frame structure.

IEEE 802.15.4 supports up to $250$ Kbps data rate with possible coverage of $10$ meters. This data rate is not enough to support the required rates of WBANs that is up to $10$ Mbps. According to  IEEE 802.15.4, packets are transmitted in the contention period, which may result longer delays in real time critical applications. When traffic is increased, the nodes compete for the contention based slots, resulting in long delays and the actual size of the network is almost doubled [7].

In [10], to satisfy the requirements of WBANs including QoS, network scalability, support for multiple PHY's and multiple application traffics, IEEE 802.15.4 MAC is proposed. It is the modified version of original IEEE 802.15.4. QoS means to decrease the packet latency and increase the probability of successful transmission of data packets without collision and loss of data. In original IEEE 802.15.4, GTS mechanism is provided to support the emergency data. GTS is very effective for data transfer, however, inherently the limit of GTS in a super frames is seven. As a result, it cannot support more than seven devices simultaneously in CFP. Whereas, in IEEE 802.15.4 MAC, the coordinator may allocate more than seven GTS simultaneously to the sensor devices.

IEEE 802.15.4 MAC and 802.15.4 original is compared briefly in Table. 5.

\section{Conclusion}
We present a survey of different MAC protocols with respect to energy efficiency and their advantages and disadvantages in WBANs. Low power listening, scheduled contention and TDMA are also compared. It is observed that TDMA is more power efficient, however, suffers with synchronization sensitivity. Techniques for collision avoidance of different MAC protocols are also comparatively analyzed. Path loss model for In-body, On-body and Off-body communication in WBANs is also described. Because human body is composed of tissues and organs in which communication is difficult and thus results in high path loss. On-body and Off-body also show some variations in results when the source and destination sensors or nodes are LoS and NLoS.


\begin{thebibliography}{00}
\bibitem{1} Anand Gopalan, S. and Park, J.T., ``Energy-efficient MAC protocols for wireless body area networks: Survey'', ICUMT, 2010.
\bibitem{2} Kutty, S. and Laxminarayan, JA., `` Towards energy efficient protocols for wireless body area networks'', ICIIS, 2010.
\bibitem{3} Ullah, S. and Shen, B. and Riazul Islam, SM and Khan, P. and Saleem, S. and Sup Kwak, K., `` A study of MAC protocols for WBANs'', SENSOR, 2009.
\bibitem{4}Omeni, O. and Wong, A. and Burdett, A.J. and Toumazou, C, ``Energy efficient medium access protocol for wireless medical body area sensornetworks", IEEE, 2008.
\bibitem{5} Timmons, NF and Scanlon, WG., ``An adaptive energy efficient MAC protocol for the medical body area network'', VITAE, 2009.
\bibitem{6} Marinkovic, S.J. and Popovici, E.M. and Spagnol, C. and Faul, S. and Marnane, W.P., `` Energy-efficient low duty cycle MAC protocol for wireless body area networks'', IEEE, 2009.
 \bibitem{7} Fang, G. and Dutkiewicz, E., ``BodyMAC: Energy efficient TDMA-based MAC protocol for wireless body area networks'',  ISCIT, 2009.
 \bibitem{8} W. Ye, J. Heidemann, and D. Estrin, ``An energy-efficient MAC protocol for wireless sensor networks'', In Proceedings of the IEEE Infocom, New York, USA, pp. 1567-1576, Jun. 2002.
 \bibitem{9} T. Van Dam and K. Langendoen, ``An adaptive energy-efficient MAC protocol for wireless sensor networks'', In ACM Conference on Embedded Networked Sensor Systems (Sensys), Los Angeles, USA, pp. 171-180, Nov. 2003.

\bibitem{10} Li, C. \textit{et al.}, ``Scalable and robust medium access control protocol in wireless body area networks'', IEEE, 2009.
\bibitem{11} Ullah, S. and Kwak, K.S., ``An ultra low-power and traffic-adaptive medium access control protocol for wireless body area network", Journal of Medical Systems, 2010.








   %
%
%
%


\end{thebibliography}
\end{document}